\documentclass[iop]{emulateapj}%[preprint2]{aastex}

\usepackage{graphicx}
\usepackage{epsfig}
\usepackage{epstopdf}
\usepackage[colorlinks=true,urlcolor=blue,citecolor=blue,linkcolor=blue]{hyperref}
\usepackage[T1]{fontenc}
\usepackage{amsmath}

\newcommand{\CaIIIR}{\ion{Ca}{2}~8542}
\newcommand{\Halpha}{H\ensuremath{\alpha}}
\newcommand{\Mgk}{\ion{Mg}{2}~k}
\newcommand{\kms}{km~s$^{-1}$}

\begin{document}

\title{Heating signatures in the disk counterparts of solar spicules in {\it IRIS} observations}

\author{L. Rouppe van der Voort\altaffilmark{1}}
\author{B. De Pontieu\altaffilmark{1,2}}
\author{T.M.D. Pereira\altaffilmark{1}}
\author{M. Carlsson\altaffilmark{1}}
\author{V. Hansteen\altaffilmark{1}}

\affil{\altaffilmark{1}Institute of Theoretical Astrophysics,
  University of Oslo, %
  P.O. Box 1029 Blindern, N-0315 Oslo, Norway}
\affil{\altaffilmark{2}Lockheed Martin Solar \& Astrophysics Lab, Org.\ A021S,
  Bldg.\ 252, 3251 Hanover Street Palo Alto, CA~94304 USA}

\begin{abstract}
We use coordinated observations with the Interface Region Imaging Spectrograph (IRIS) and the Swedish 1-m Solar Telescope (SST) to identify the disk counterpart of type II spicules in upper-chromospheric and transition region (TR) diagnostics.
These disk counterparts were earlier identified through short-lived asymmetries in chromospheric spectral lines: rapid blue- or red-shifted excursions (RBEs or RREs). 
We find clear signatures of RBEs and RREs in \ion{Mg}{2}~h \& k, often with excursions of the central h3 and k3 absorption features in concert with asymmetries in co-temporal and co-spatial \Halpha\ spectral profiles. 
We find spectral signatures for RBEs and RREs in \ion{C}{2}~1335 and 1336~\AA\ and \ion{Si}{4}~1394 and 1403~\AA\ spectral lines and interpret this as a sign that type II spicules are heated to at least TR temperatures, supporting other recent work.
These \ion{C}{2} and \ion{Si}{4} spectral signals are weaker for a smaller network region than for more extended network regions in our data.
A number of bright features around extended network regions observed in IRIS slit-jaw imagery SJI 1330 and 1400, recently identified as network jets, can be clearly connected to \Halpha\ RBEs and/or RREs in our coordinated data. 
We speculate that at least part of the diffuse halo around network regions in the IRIS SJI 1330 and 1400 images can be attributed to type II spicules with insufficient opacity in the \ion{C}{2} and \ion{Si}{4} lines to stand out as single features in these passbands. 
\end{abstract} 
% Note: max 250 words

% 3500 words in main body
% 5 Figures + Tables, max 9 panels per figure

\section{Introduction}
\label{sec:intro}

Spicules are highly dynamic, linear features that 
can be observed as jet-like extrusions that emanate from the solar limb
\citep[for a recent review see][]{2012SSRv..tmp...65T}. %  Tsiropoula review
Two classes of spicules have been identified
of which the second class is the most energetic displaying shortest lifetimes and most vigorous dynamical evolution
\citep{2007PASJ...59S.655D, % BdP Tale of 2 spicules
2012ApJ...759...18P}. % Tiago: Quantifying
The complex dynamics of these type II spicules has been identified as the interplay between three kinds of motion: up flow, transversal, and torsional motions 
\citep{2012ApJ...752L..12D}. % BdP: type II torsional motions
A recent study combing observations from Hinode and the Interface Region Imaging Spectrograph (IRIS) showed that type II spicules are heated to transition region (TR) temperatures
\citep{2014ApJ...792L..15P}. % Tiago 2014 IRIS spicules
This strengthens earlier observational indications that type II spicules are heated to coronal temperatures
\citep{2011Sci...331...55D}. % BdP: Origins of hot plasma

The true impact of spicules on the corona in terms of provision of mass and energy is still unclear and a topic of debate. 
A prerequisite to realistic modelling of spicules is detailed knowledge of their physical properties. 
A full observational characterisation is, however, hampered by line-of-sight confusion due to superposition effects at the limb.
This problem is relieved on the solar disk where single spicules can be identified unambiguously. 
The disk counterparts of type II spicules have been identified through short-lived asymmetries of chromospheric spectral lines and are referred to as rapid blue- or red-shifted excursions 
\citep[RBEs or RREs,][]{2008ApJ...679L.167L, % Langangen
2009ApJ...705..272R, % Rouppe: RBE
2012ApJ...752..108S, % Sekse: statistical RBE
2013ApJ...764..164S, % Sekse: temporal RBE
2013ApJ...769...44S, % Sekse: RRE
2013ApJ...767...17Y}. % Yurchyshyn: NST RBEs

In this study we use coordinated observations from IRIS and the Swedish 1-m Solar Telescope (SST). 
RBEs and RREs are identified in the SST data and we search for corresponding signal in various IRIS diagnostics.

%------------------------------------------------------------
\section{Observations and Data Processing}
\label{sec:obs}

We study Quiet Sun and Coronal Hole observations from a coordinated IRIS / SST observing campaign during the period 25-August -- 3-Oct, 2013. 
IRIS observes spectra and slit-jaw images in a number of spectral windows in the near- and far-UV 
\citep[for details see][]{2014SoPh..289.2733D}. %BDP IRIS
Here we concentrate on spectral diagnostics of the transition region:  \ion{Si}{4}~1394 and 1403~\AA, the upper chromosphere/transition region: \ion{C}{2}~1335 and 1336~\AA, and the chromosphere: \ion{Mg}{2} h \& k (2803 and 2796~\AA). 
Spectra were acquired with the IRIS slit at a fixed solar position (following solar rotation, a so-called ``sit-and-stare'' program), or with narrow spatial rasters, Table~\ref{tab:datasets} provides details for the different datasets. 
Besides the spectra, we analyse IRIS slit-jaw images (commonly referred to as SJI): SJI 1400 (dominated by the \ion{Si}{4} lines), SJI 1330 (dominated by the \ion{C}{2} lines), and SJI 2796 (\Mgk\ core and inner wings).
Exposure times for both spectra and slit-jaw images were 4~s, the pixel size 0\farcs166. 
The IRIS observations were processed to level3 data
\citep[for more details, see][]{2014SoPh..289.2733D}. %BDP IRIS

With the Crisp Imaging Spectropolarimeter 
\citep[CRISP][]{2008ApJ...689L..69S} % CRISP
at the SST
\citep{2003SPIE.4853..341S}, %SST
we acquired spectra in \Halpha\ and \CaIIIR, and Stokes~V maps in \ion{Fe}{1}~6302 at $-48$~m\AA\ at a temporal cadence of about 11~s.
\Halpha\ was symmetrically sampled at 15 line positions, with 200~m\AA\ steps, \CaIIIR\ at 25 line positions, with 100~m\AA\ steps. 
High spatial resolution was achieved with the aid of the adaptive optics system (recently upgraded with an 85-electrode deformable mirror), and with image restoration using the Multi-Object Multi-Frame Blind Deconvolution 
\citep[MOMFBD,][]{2005SoPh..228..191V} % 
method.
The CRISP data reduction pipeline
\citep{2014arXiv1406.0202D} % Jaime CRISPRED
includes different methods described in detail by
\citet{2013A&A...556A.115D, % Jaime 8542 backscatter
2012A&A...548A.114H} %Henriques post-MOMFBD destretch
and \citet{2008A&A...489..429V}. % van Noort & Rouppe : polarimetry with MOMFBD
% Shine 1994?
After data reduction, the effective field of view (FOV) of the CRISP observations is approximately $55\arcsec \times 55\arcsec$, with a pixel scale of 0\farcs058.

Alignment of the SST and IRIS data was done by scaling down the CRISP data to IRIS pixel scale and by cross-correlation of \CaIIIR\ far wing images with IRIS SJI 2796 images. 
This alignment procedure proved to be accurate down to the level of the IRIS pixel scale. 
Details of the overlap of SST and IRIS data, in time and spatially with the IRIS slit and SJI channels, %save words
are given in the last three columns of Table~\ref{tab:datasets}.
For exploration of the aligned SST and IRIS data, we used CRISPEX 
\citep{2012ApJ...750...22V}, % Gregal CRISPEX
initially developed for CRISP data and now expanded to browse IRIS level3 data.
CRISPEX is part of SolarSoft.

%----------------------------
\begin{table*}[bth]
\caption{Overview of the IRIS data sets analyzed in this study.}
\begin{center}
\begin{tabular}{cccccccc}
	\hline \hline
& & & & & \multicolumn{3}{c}{overlap with SST$^{\rm{c}}$} \\
 Date & Time (UT) & Type$^{\rm{a}}$ & FOV$^{\rm{b}}$ & Pointing & Time & SJI [arcsec$^2$]$^{\rm{d}}$ & Slit \\
 	\hline
 13-Sep-2013 & 08:17 -- 14:54 & 4-step dense raster & 1\arcsec $\times$ 50\arcsec & Quiet Sun disk center & 02:13 & 2366 (95\%) & 43\arcsec \\
 18-Sep-2013 & 08:00 -- 11:30 & 2-step sparse raster & 1\arcsec $\times$ 50\arcsec & Quiet Sun (37\arcsec, 58\arcsec) & 00:11 & 2204 (88\%) & 42\arcsec \\
 22-Sep-2013 & 07:34 -- 11:04 & medium sit-and-stare & 0\farcs33 $\times$ 61\arcsec & Coronal Hole (538\arcsec, 283\arcsec) & 02:03 & 1372 (38\%) & 26\arcsec \\
 23-Sep-2013 & 07:09 -- 12:05 & medium sit-and-stare & 0\farcs33 $\times$ 61\arcsec & Quiet Sun disk center & 00:46 & 1323 (37\%) & 15\arcsec \\
  	\hline
\end{tabular}
\begin{minipage}{.9\hsize}
  $^{\rm{a}}$ A dense raster has 0\farcs33 slit steps, a sparse raster has 1\arcsec\ steps. A sit-and-stare program keeps the slit at a fixed position. \\
  $^{\rm{b}}$ Area covered by the spectrograph slit. \\
  $^{\rm{c}}$ The last three columns provide information on the overlap of the SST and IRIS data. \\
  $^{\rm{d}}$ The overlap of SST was measured against SJI 2796 (percentage of SJI pixels covered in the SST FOV).
\end{minipage}
\end{center}
\label{tab:datasets}
\end{table*}%
%% overlap SST time, overlap SST / SJI, overlap SST / slit, 

%-----------------------------------------
\section{Results}
\label{sec:results}

\begin{figure*}[!t]
\begin{center}
\includegraphics[width=\textwidth]{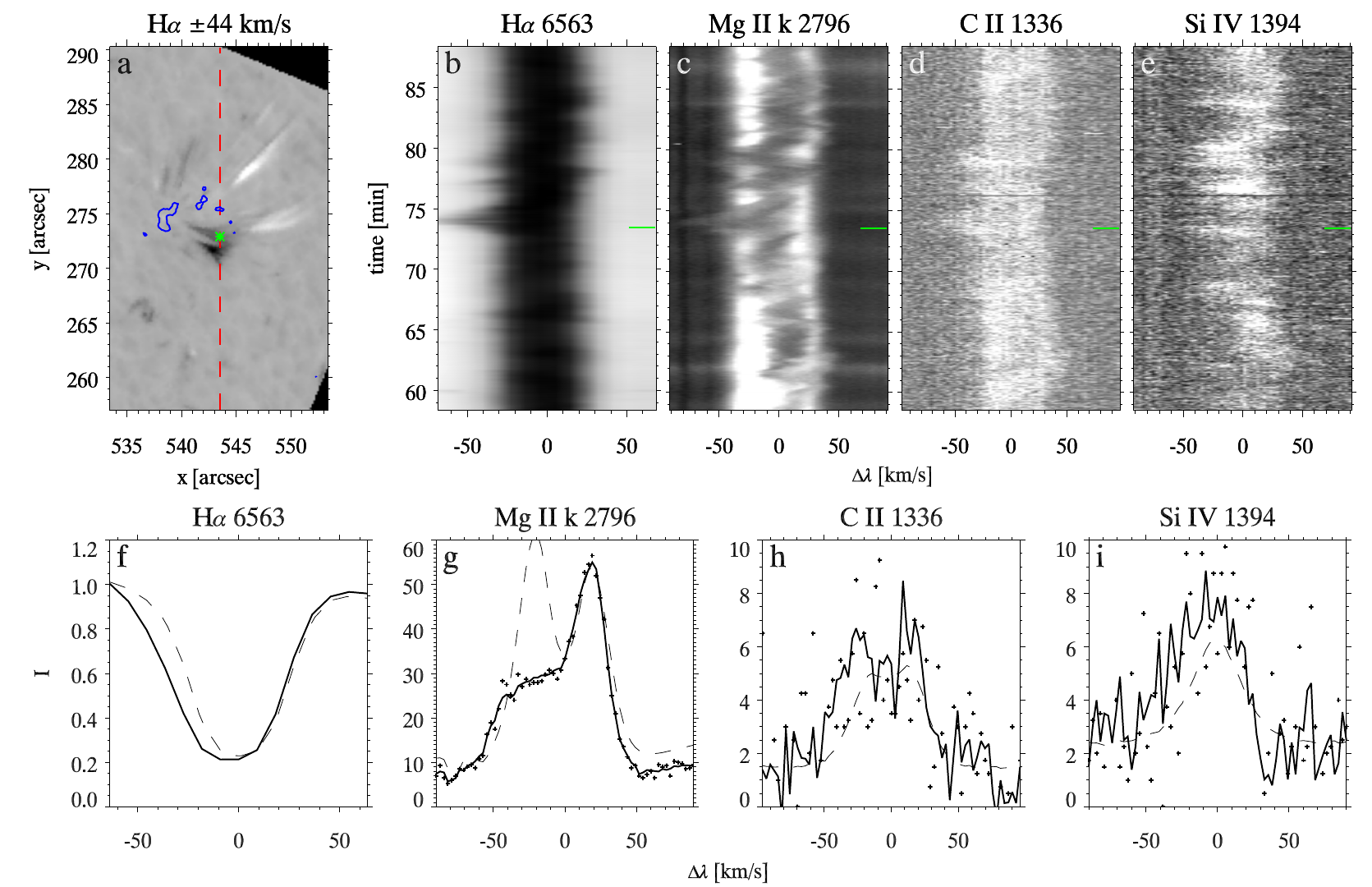}
\caption{Example of an RBE under the IRIS slit. Panel a shows a CRISP \Halpha\ Dopplergram at $\pm44$~\kms, RBEs are black, RREs are white. The location of the IRIS slit is indicated with the red dashed line. The blue contour outlines the strongest signal in the \ion{Fe}{1} Stokes V map (negative polarity). Panels b--e show $\lambda t$-diagrams of the location marked with the green asterisk in panel a, where the slit covers the RBE. The data for the IRIS diagnostics in c--e are averaged over 3 neighbouring spatial locations to reduce noise. The green line marks the time of the RBE in panel a and for which spectral line profiles are shown in panels f--i. The thick solid line in panels f--i is the RBE spectral profile, for the IRIS lines (g--i) averaged over 3 spatial locations and 3 time steps (i.e., an average of 9 spectra), the small crosses mark the data points of one spectrum (i.e., one spatial pixel) centred on the RBE event in panel a. The thin dashed line is a reference spectrum, spatially and temporally averaged over slit positions far away from the network region. A movie with the \Halpha\ Dopplergrams is available in the on-line material.}
\label{fig:rbe}
\end{center}
\end{figure*}

\begin{figure*}[!t]
\begin{center}
\includegraphics[width=\textwidth]{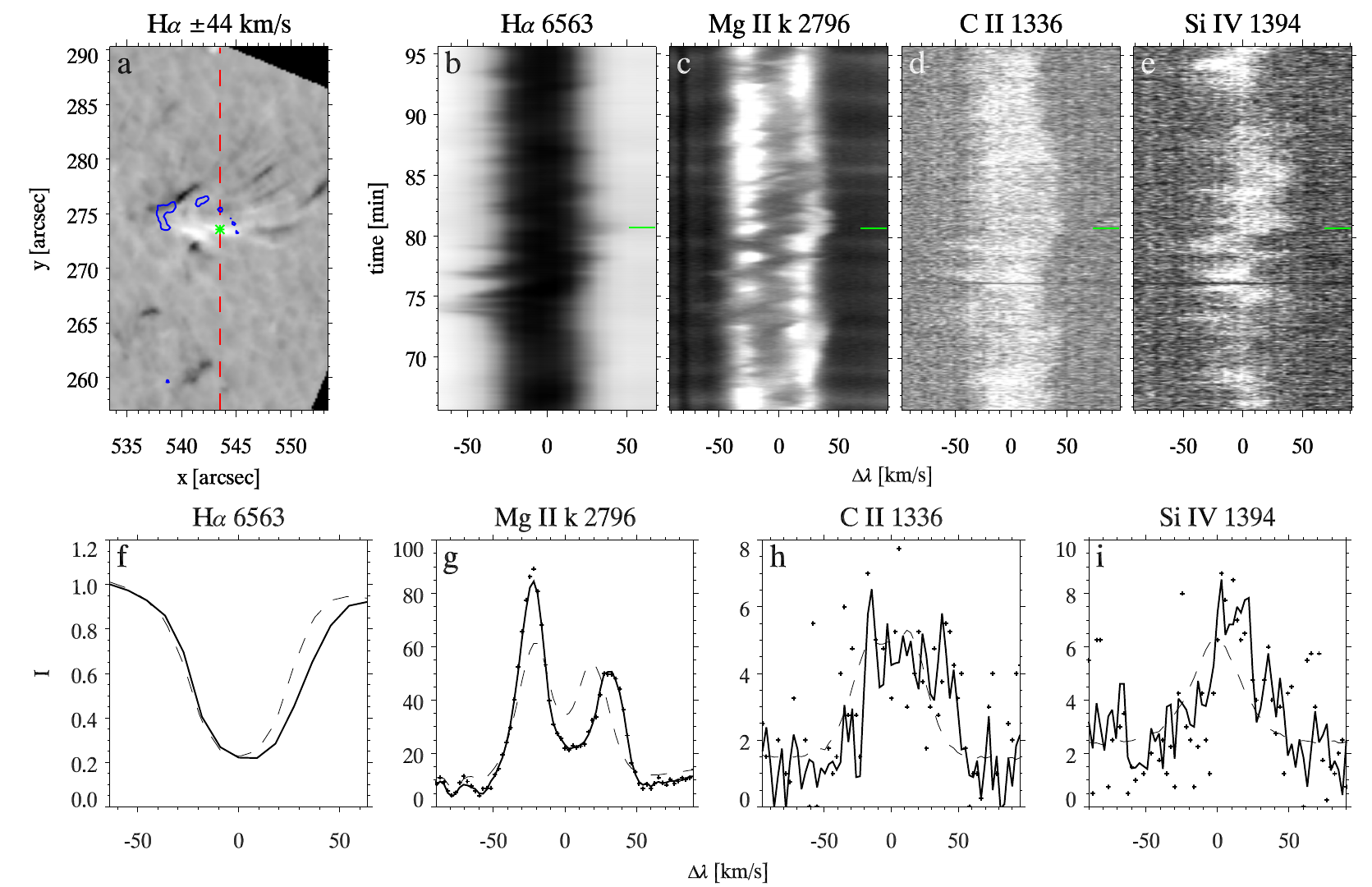}
\caption{Example of an RRE under the IRIS slit. The panel layout is the same as in Fig.~\ref{fig:rbe}.}
\label{fig:rre}
\end{center}
\end{figure*}

RBEs and RREs are most easily recognised in \Halpha\ Doppler maps (blue $-$ red far wing subtraction images) and are most frequently found around network regions.
In the 22-Sep-2013 dataset, the IRIS slit was positioned in the vicinity of a small network patch 
and we observe a large number 
(more than 150 in 2~h) %L I measured 182 in February
%($>150$ in 2~h) %L I measured 182 in February
of RBE/RREs that are covered by the IRIS slit.
This is illustrated in \Halpha\ $\pm44$~\kms\ Doppler maps in Figs.~\ref{fig:rbe} and \ref{fig:rre} panels $a$ and the accompanying movie: RBEs (dark streaks) and RREs (white) appear to originate radially from the region around $(x,y) \approx (538\arcsec,275\arcsec)$ and cross the IRIS slit (red dashed line). 
The spectral evolution ($\lambda t$-diagrams) of one location at the slit (marked with a green asterisk),
 is shown in panels $b$--$e$.
The \Halpha\ $\lambda t$-diagram ($b$) shows the occurrence of a large number of RBEs and RREs at this location; short-lived asymmetry excursions of the line, most prominently the RBE around $t=74$~min, the time of the Dopplergram in Fig.~\ref{fig:rbe}$a$ which shows the RBE as a dark streak crossed by the IRIS slit at the green asterisk. 
With the blue- and red-shifted excursions in the \Halpha\ $\lambda t$-diagram at hand, the RBEs and RREs can be readily identified as similar excursions of the central absorption feature k3 in the \Mgk\ diagram ($c$). 
The shift of k3 leads to enhancement of the wing 
so that the RBE/RREs appear as emission features far out in the wings.
The line profile in panel \ref{fig:rbe}$g$ shows the RBE as a blue-shift of \ion{Mg}{2}~k3 and absence of a prominent k2v peak as compared to the average \Mgk\ profile (constructed from temporal and spatial averaging over slit positions far away from the network region).  
For the RRE example in Fig.~\ref{fig:rre}, the k2r peak is weakened but not absent, and shifted with more than 10~\kms\ as compared to the average profile. 
Even more than the RBE example in Fig.~\ref{fig:rbe}, this RRE appears as an emission feature extending out to $\sim+40$~\kms\ in the $\lambda t$-diagram ($c$).

The far-UV spectra for \ion{C}{2}~1336~\AA\ and \ion{Si}{4}~1394~\AA\ are very noisy at this exposure time. 
However, comparing panels $d$ and $e$ with $b$ and $c$, these lines show similar temporal behaviour with blue- and red-ward asymmetry excursions in concert with their pure chromospheric counterparts \Halpha\ and \Mgk.
After a 9-fold averaging (3 slit positions over 3 time steps), the line profiles (solid lines in panels $h$ and $i$) show asymmetry (and shift) towards the blue for the RBE and towards the red for the RRE. 
Single Gaussian fits to the UV spectral profiles indicate shifts in the range 11--14~\kms\ for both the RBE and RRE in Figs.~\ref{fig:rbe} and \ref{fig:rre}.  

As the signals in these far-UV IRIS diagnostics are low, the associated slit-jaw images SJI 1330 and SJI 1400 show no sign of anything related to the RBEs and RREs that are so prominent in the CRISP data in this region. 
This network region can be considered to be weak in a sense that it is only a collection of a small number of photospheric magnetic bright points.  
However, in other more extended network regions with larger numbers of bright points, the slit jaw images show a pattern of a diffuse halo around the network patch, frequently interspersed with dynamic bright streaks that often appear to move away from the network region. 
These streaks are recently described as network jets by 
\citet{tian2014science}. % Tian 2014 Science
Many of these network jets 
can be directly related to RBEs and RREs in the CRISP Doppler movies: they can often be associated with an RBE or RRE in close vicinity with similar orientation and temporal evolution (life time and direction of propagation). 

\begin{figure*}[!t]
\begin{center}
\includegraphics[width=\textwidth]{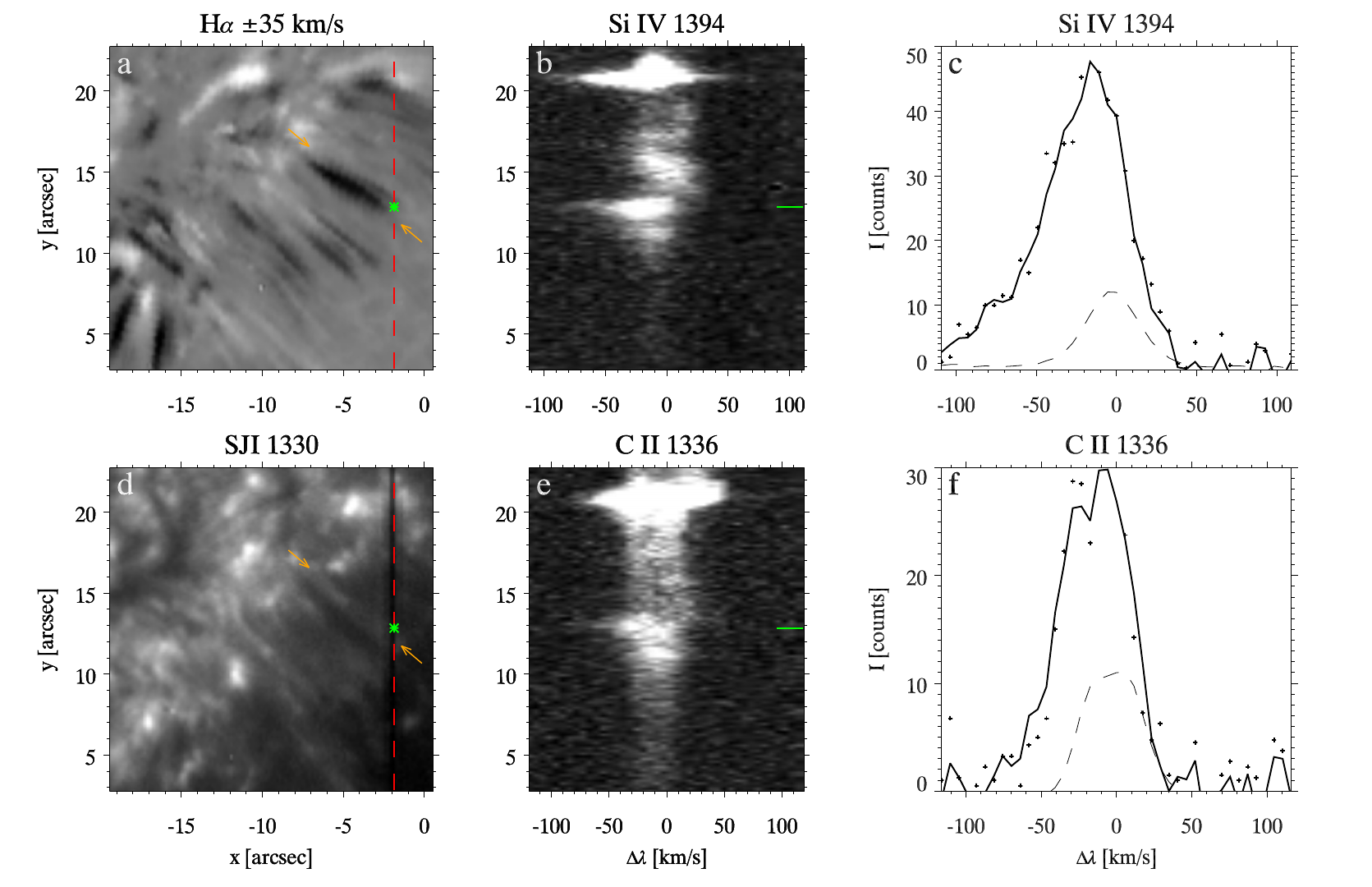}
\includegraphics[width=\textwidth]{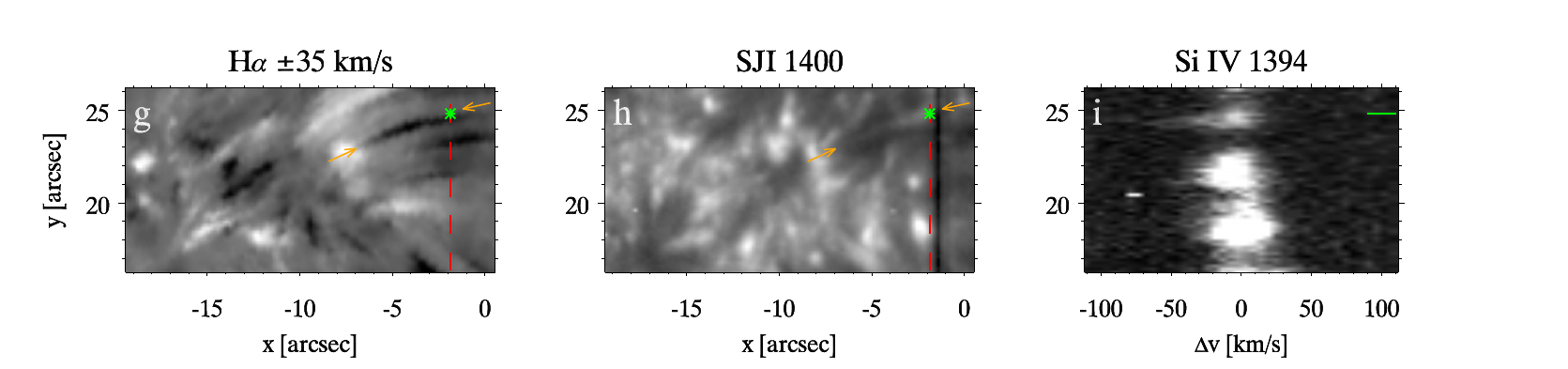}
\caption{Examples of SJI jets associated with \Halpha\ RBEs. Panels a--f show details of one event, panels g--i of another event. The vertical red dashed line in panels a, d, g, and h, marks the location of the IRIS slit used for the data in the other panels. Orange arrows mark the approximate end points of the events. Panels b, e, and i show $\lambda y$-diagrams. The green horizontal lines in these panels mark the spatial location of the green asterisk in the associated images. The thick solid lines in panels c and f are spectral line profiles averaged over 9 spectra. Small crosses are data points from the spectrum marked with the green line in panels b and e. The thin dashed lines are reference spectra. }
\label{fig:jets}
\end{center}
\end{figure*}

Figure~\ref{fig:jets} shows two examples of far-UV slit jaw jets 
in the 13-Sep-2013 Quiet Sun data set. 
Panel $d$ shows a SJI 1330 jet, 
%and 
panel $a$ shows the associated \Halpha\ RBE in the co-temporal CRISP Doppler map, and
panels $b$ and $e$ show the co-temporal spectrograms ($\lambda y$-diagrams). 
In contrast with the spectra in Figures~\ref{fig:rbe} and \ref{fig:rre}, the \ion{Si}{4}~1394~\AA\ and \ion{C}{2}~1336~\AA\ spectra show strong signal with clearly shifted and blue-ward asymmetric line profiles.
Gaussian fits to these line profiles indicate blue shifts of $-16$~\kms\ for \ion{Si}{4} and $-12$~\kms\ for \ion{C}{2} with respect to the average line profiles. We note the asymmetry of the profiles towards higher Doppler shift that is not captured by a single Gaussian profile.

The bottom three panels in Fig.~\ref{fig:jets} show another example: a bright SJI 1400 jet
($h$), associated with an \Halpha\ RBE in $g$ and a strong asymmetric profile in the \ion{Si}{4}~1394~\AA\ spectrogram.

\begin{figure*}[!t]
\begin{center}
\includegraphics[width=\columnwidth]{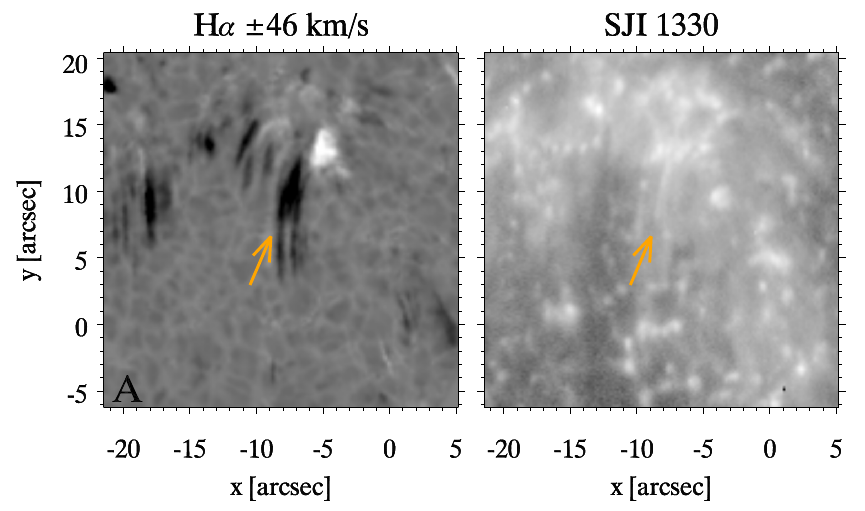} %A
\includegraphics[width=\columnwidth]{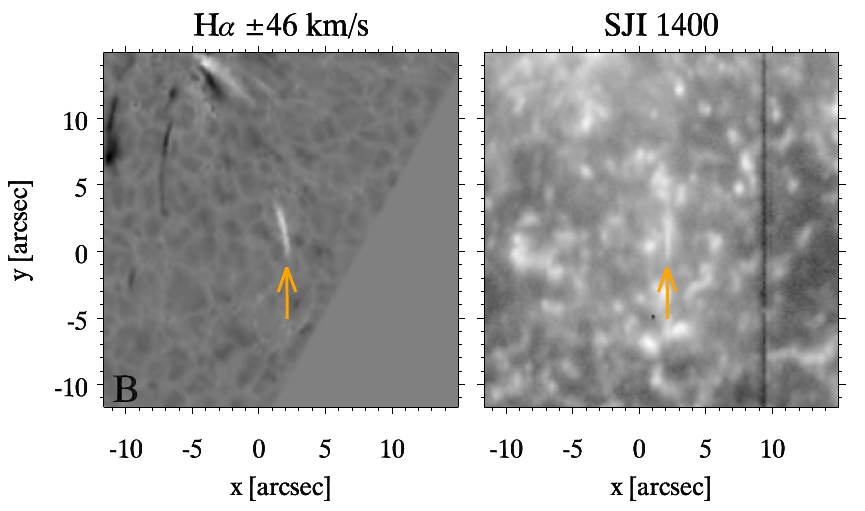} %B
\includegraphics[width=\columnwidth]{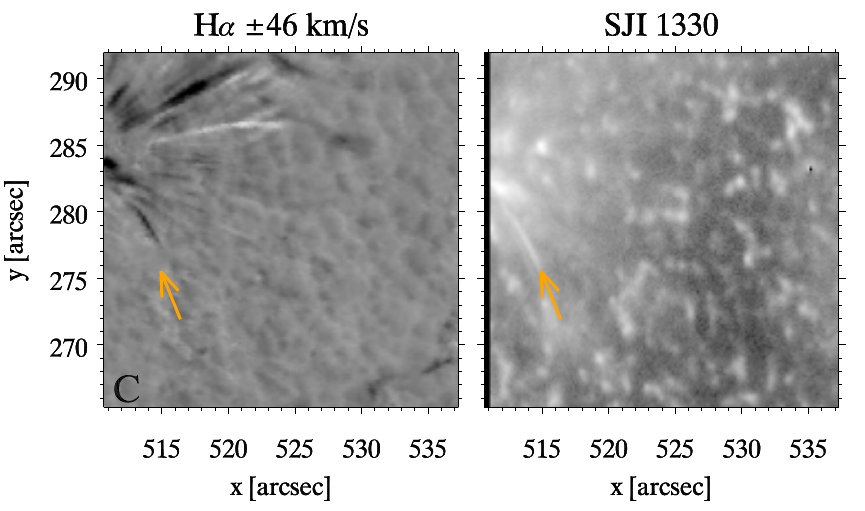} %C
\includegraphics[width=\columnwidth]{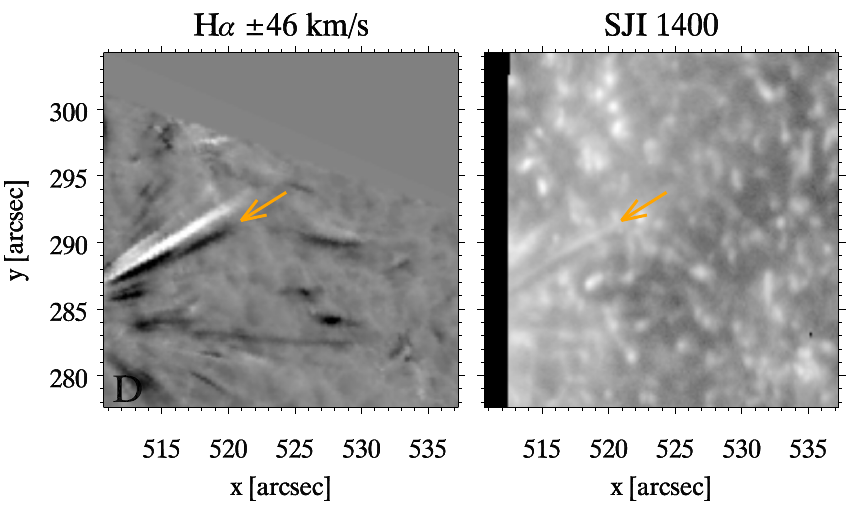} %D
\caption{Four examples of SJI jets associated with \Halpha\ RBEs or RREs. Orange arrows guide the eye to identify the events in both images for each pair. Movies for each event are available in the online material. }
\label{fig:sjijets}
\end{center}
\end{figure*}

Figure~\ref{fig:sjijets} shows four more examples of jets %streaks 
in SJI 1330 or SJI 1400 associated with RBEs or RREs in \Halpha\ Doppler maps. 
The features of interest are marked with arrows in the different panels, but the association between IRIS SJI jets %streaks 
and CRISP RBE/RREs is most vivid in the accompanying movies available as online material. 

\section{Discussion}
\label{sec:discussion}

We use coordinated observations from the SST and IRIS to identify the disk counterparts of type II spicules in various IRIS chromospheric and TR diagnostics. 
One of the data sets is a more than 2~h time series with the IRIS spectrograph slit positioned in the proximity ($\sim5\arcsec$) of a small network region in a coronal hole. 
A large number of \Halpha\ RBEs and RREs ($>$150) were found to be covered by the IRIS slit.
For these events, we find clear correspondence between the temporal behaviour of the \Halpha\ line and the \ion{Mg}{2}~h \& k lines: for \Halpha\ RBEs we find corresponding Mg blue-shifted spectral features and for RREs we find corresponding red-shifted features. 
We observe that the k3 and h3 central absorption dips show similar Doppler excursions as the asymmetries in the \Halpha\ line. 
This often leads to a strong reduction (sometimes almost complete removal) of one of the k2 and h2 reversal peaks accompanied with a Doppler shift of this peak. 
The shift results in enhanced emission in the wing so that RBE/RREs can be observed as bright features in the \ion{Mg}{2}~h \& k wings.
This is different from \Halpha\ and \ion{Ca}{2}~8542~\AA\ where RBE/RREs are pure absorption features and therefore appear dark in wing images.

With 4~s exposure times, the \ion{C}{2} and \ion{Si}{4} spectra in this coronal hole region are to a large extent dominated by noise. 
However, after averaging of spectra (both spatially and temporally), we are able to identify spectral signatures for these lines that can be associated with \Halpha\ and Mg RBEs and RREs. 
We often see (weak) enhanced emission with corresponding Doppler shift for both the \ion{C}{2} and \ion{Si}{4} lines. 
In a different data set, with the IRIS slit positioned in the proximity of a more extended magnetic network region in a Quiet Sun region, we observe for some \Halpha\ RBEs and RREs much stronger response in these upper-chromospheric and TR diagnostics. 
For these events, no spectral averaging is needed and the presence of \ion{C}{2} and \ion{Si}{4} signal corresponding to \Halpha\ RBEs or RREs is unquestionable.  

The RBEs and RREs that have strong \ion{C}{2} and \ion{Si}{4} response can be clearly identified as bright streaks in the SJI 1330 and 1400 slit-jaw images. 
These kind of bright streaks can be found around the stronger network regions anywhere on the solar disk in the IRIS SJI 1330 and 1400 data,
\citet{tian2014science} % Tian 2014 Science
describe these features as network jets.
When we have coordinated SST \Halpha\ observations, we can often identify corresponding RBEs and RREs that display striking similarity in morphology and dynamical evolution.
This suggests that we observe the same feature in different diagnostics. %Bart: split
Here we present 4 examples of \Halpha\ RBE/RREs that can be associated with SJI 1330 or 1400 jets. 
% 3 events in Science paper
\citet{2014Sci...346D.315D} %BdP 2014 Science twist % depontieu2014science} 
present 3 other events from the same data sets for which SJI 1330 and 1400 jets are associated with RBE/RREs that display clear twist in their dynamical evolution. 
The relatively broad bandpasses for the SJI 1330 and 1400 filters  (55~\AA\ for both) cover a large continuum window. The strong line emission we observe for jets that are covered by the IRIS slit indicates that the network jets in the slit jaw data are dominated by line emission rather than continuum contribution.

We conclude that our observations show that type II spicules are heated to at least TR temperatures, interpreting the \ion{Si}{4} emission as signs of $\sim80$~kK plasma (under equilibrium conditions). 
This is in agreement with 
\citet{2014ApJ...792L..15P} % Tiago 2014 IRIS spicules
who tied Hinode \ion{Ca}{2}~H type II spicules at the Quiet Sun limb to IRIS SJI 1400 spicules. 
Their study demonstrates that the Hinode Ca~H passband shows type II spicules only during their initial (cooler) phase and that the spicules continue to evolve in the hotter SJI 1400 passband. 
Furthermore, \citet{2014ApJ...792L..15P} % Tiago 2014 IRIS spicules
identified RBE and RRE signatures in \ion{Mg}{2}~k spectroheliograms from dense IRIS rasters close to the limb.
Many of these Mg RBEs and RREs could be seen to continue in \ion{Si}{4} spectroheliograms.
Our study supports these results and further establish a close connection between RBE/RREs observed in \Halpha\ and \ion{Mg}{2}~h \& k. 

We note the difference between the small network region close to the IRIS slit in the 22-Sep-2013 coronal hole data set and the more extended network regions in the same and other data sets. For the small network region we only find weak \ion{C}{2} and \ion{Si}{4} signal and an absence of network jets in SJI 1330 and 1400. For the more extended network regions we find network jets and associated features in the spectrograms. In the CRISP \Halpha\ and \ion{Ca}{2}~8542 data however, we cannot identify a clear difference in RBE/RRE activity between these network regions. 
Apparently, while these network regions seem indistinguishable in chromospheric diagnostics, for the smaller network region insufficient opacity is built up in \ion{C}{2} and \ion{Si}{4}.
Whether this is due to a different level of heating or mass loading can only be speculated on.

The network jets appear often to be embedded in diffuse halos that surround the larger network regions in SJI 1330 and 1400 data, while these halos appear to be absent in small network regions like in the 22-Sep-2013 data set. 
Apparently, the weak \ion{C}{2} and \ion{Si}{4} emission that we effectively find for almost all RBEs and RREs is insufficient to give discernible signal in the slit jaw data around small network regions.
It might however be well possible that the diffuse halo around the larger network regions result from type II spicules with insufficient opacity to stand out as individual network jets in the slit jaw images.

We find a number of clear examples of SJI 1330 and 1400 network jets that are connected to \Halpha\ RBEs and/or RREs. 
However, we cannot firmly establish a chromospheric RBE/RRE connection for all network jets in our data.
It is possible that this is due to a line-of-sight effect: some network jets might follow a dynamic trajectory that is unfavourable for identification in \Halpha\ wing Dopplergrams. 
Furthermore, it cannot be excluded that some network jets simply do not have a chromospheric component or are observed during a more energetic part of their evolution that makes it difficult to connect to a possible earlier chromospheric phase.
\citet{tian2014science} connect network jets to the solar wind and measure apparent velocities that are often well above 100~\kms. 
These apparent velocities are on the high end of
what has been found earlier for RBE/RREs and type II spicules.
The Doppler shifts we measure for \ion{C}{2} and \ion{Si}{4} are in the low range of what was found for RBE/RREs 
\citep{2013ApJ...764..164S}. % Sekse temporal
We note, however, that single Gaussian fits do not catch the full complexity of these profiles. For example, there appears to be an extra component at $\sim-75$~\kms\ in the far blue wing of the \ion{Si}{4} profile in Fig.~\ref{fig:jets}c.
Further effort is required to explore how the asymmetry in the profiles can be characterised.
This is crucial in order to determine to what extent the high apparent speeds in network jets are real mass flows.

In this Letter, we do not address the temporal evolution of the spicules in our coordinated data set. We note, however, the prominent appearance of swings between red and blue shifts in the \ion{Mg}{2} $\lambda t$-diagrams. It seems that this temporal variation is much more clearly tractable than in the ground-based \Halpha\ and \CaIIIR\ data. Potentially, this opens a new diagnostic window on the Alfv{\'e}nic wave modes that govern the dynamics of spicules. 
However, disentangling the observed Doppler shifts in uniquely determined kink, torsional and up-flow components will remain to be a formidable challenge.

\acknowledgements
We thank 
J. De la Cruz-Rodr{\'i}guez, 
T. Golding, 
C. Hoffman, 
S. Jafarzadeh, 
L. Kleint, 
A. Ortiz, 
E. Scullion, 
T. Tarbell, 
A. Sainz-Dalda, 
and H. Skogsrud
for assistance with SST observations and data processing. 
IRIS is a NASA Small Explorer mission developed and operated by LMSAL with mission operations executed at NASA ARC and major contributions to downlink communications funded by the NSC (Norway).
The Swedish 1-m Solar Telescope is operated on the island of La Palma by the Institute for Solar Physics (ISP) of Stockholm University in the Spanish Observatorio del Roque de los Muchachos of the Instituto de Astrof\' isica de Canarias.
Part of the CRISP data processing was performed on computing resources
kindly provided by the ISP. % Institute for Solar Physics of Stockholm University. % ISP, define above
The authors gratefully acknowledge support from the International Space Science Institute (ISSI).
This research was supported by the Research Council of Norway and by the European Research Council under the European Union's Seventh Framework Programme (FP7/2007-2013) / ERC Grant agreement nr. 291058.
% Bart:
B.D.P. is supported by NASA
contract NNG09FA40C (IRIS), and NASA grants NNX11AN98G and NNM12AB40P.
This research has made use of NASA's Astrophysics Data System.

%\bibliographystyle{as}
%\bibliographystyle{apj}
%\bibliography{irisrbe_references}

\begin{thebibliography}{23}
\expandafter\ifx\csname natexlab\endcsname\relax\def\natexlab#1{#1}\fi

\bibitem[{{de la Cruz Rodr{\'{\i}}guez} {et~al.}(2015){de la Cruz
  Rodr{\'{\i}}guez}, {L{\"o}fdahl}, {S{\"u}tterlin}, {Hillberg}, \& {Rouppe van
  der Voort}}]{2014arXiv1406.0202D}
{de la Cruz Rodr{\'{\i}}guez}, J., {L{\"o}fdahl}, M., {S{\"u}tterlin}, P.,
  {Hillberg}, T., \& {Rouppe van der Voort}, L. 2015, \aap, 573, A40

\bibitem[{{de la Cruz Rodr{\'{\i}}guez} {et~al.}(2013){de la Cruz
  Rodr{\'{\i}}guez}, {Rouppe van der Voort}, {Socas-Navarro}, \& {van
  Noort}}]{2013A&A...556A.115D}
{de la Cruz Rodr{\'{\i}}guez}, J., {Rouppe van der Voort}, L., {Socas-Navarro},
  H., \& {van Noort}, M. 2013, \aap, 556, A115

\bibitem[{{De Pontieu} {et~al.}(2012){De Pontieu}, {Carlsson}, {Rouppe van der
  Voort}, {Rutten}, {Hansteen}, \& {Watanabe}}]{2012ApJ...752L..12D}
{De Pontieu}, B., {Carlsson}, M., {Rouppe van der Voort}, L.~H.~M., {Rutten},
  R.~J., {Hansteen}, V.~H., \& {Watanabe}, H. 2012, \apjl, 752, L12

\bibitem[{{De Pontieu} {et~al.}(2007){De Pontieu}, {McIntosh}, {Hansteen},
  {Carlsson}, {Schrijver}, {Tarbell}, {Title}, {Shine}, {Suematsu}, {Tsuneta},
  {Katsukawa}, {Ichimoto}, {Shimizu}, \& {Nagata}}]{2007PASJ...59S.655D}
{De Pontieu}, B., {et~al.} 2007, \pasj, 59, 655

\bibitem[{{De Pontieu} {et~al.}(2011){De Pontieu}, {McIntosh}, {Carlsson},
  {Hansteen}, {Tarbell}, {Boerner}, {Martinez-Sykora}, {Schrijver}, \&
  {Title}}]{2011Sci...331...55D}
---. 2011, Science, 331, 55

\bibitem[{{De Pontieu} {et~al.}(2014{\natexlab{a}}){De Pontieu}, {Rouppe van
  der Voort}, {McIntosh}, {Pereira}, {Carlsson}, {Hansteen}, {Skogsrud},
  {Lemen}, {Title}, {Boerner}, {Hurlburt}, {Tarbell}, {Wuelser}, {De Luca},
  {Golub}, {McKillop}, {Reeves}, {Saar}, {Testa}, {Tian}, {Kankelborg},
  {Jaeggli}, {Kleint}, \& {Martinez-Sykora}}]{2014Sci...346D.315D}
---. 2014{\natexlab{a}}, Science, 346, D315

\bibitem[{{De Pontieu} {et~al.}(2014{\natexlab{b}}){De Pontieu}, {Title},
  {Lemen}, {Kushner}, {Akin}, {Allard}, {Berger}, {Boerner}, {Cheung}, {Chou},
  {Drake}, {Duncan}, {Freeland}, {Heyman}, {Hoffman}, {Hurlburt}, {Lindgren},
  {Mathur}, {Rehse}, {Sabolish}, {Seguin}, {Schrijver}, {Tarbell},
  {W{\"u}lser}, {Wolfson}, {Yanari}, {Mudge}, {Nguyen-Phuc}, {Timmons}, {van
  Bezooijen}, {Weingrod}, {Brookner}, {Butcher}, {Dougherty}, {Eder},
  {Knagenhjelm}, {Larsen}, {Mansir}, {Phan}, {Boyle}, {Cheimets}, {DeLuca},
  {Golub}, {Gates}, {Hertz}, {McKillop}, {Park}, {Perry}, {Podgorski},
  {Reeves}, {Saar}, {Testa}, {Tian}, {Weber}, {Dunn}, {Eccles}, {Jaeggli},
  {Kankelborg}, {Mashburn}, {Pust}, {Springer}, {Carvalho}, {Kleint}, {Marmie},
  {Mazmanian}, {Pereira}, {Sawyer}, {Strong}, {Worden}, {Carlsson}, {Hansteen},
  {Leenaarts}, {Wiesmann}, {Aloise}, {Chu}, {Bush}, {Scherrer}, {Brekke},
  {Martinez-Sykora}, {Lites}, {McIntosh}, {Uitenbroek}, {Okamoto}, {Gummin},
  {Auker}, {Jerram}, {Pool}, \& {Waltham}}]{2014SoPh..289.2733D}
---. 2014{\natexlab{b}}, \solphys, 289, 2733

\bibitem[{{Henriques}(2012)}]{2012A&A...548A.114H}
{Henriques}, V.~M.~J. 2012, \aap, 548, A114

\bibitem[{{Langangen} {et~al.}(2008){Langangen}, {De Pontieu}, {Carlsson},
  {Hansteen}, {Cauzzi}, \& {Reardon}}]{2008ApJ...679L.167L}
{Langangen}, {\O}., {De Pontieu}, B., {Carlsson}, M., {Hansteen}, V.~H.,
  {Cauzzi}, G., \& {Reardon}, K. 2008, \apjl, 679, L167

\bibitem[{{Pereira} {et~al.}(2012){Pereira}, {De Pontieu}, \&
  {Carlsson}}]{2012ApJ...759...18P}
{Pereira}, T.~M.~D., {De Pontieu}, B., \& {Carlsson}, M. 2012, \apj, 759, 18

\bibitem[{{Pereira} {et~al.}(2014){Pereira}, {De Pontieu}, {Carlsson},
  {Hansteen}, {Tarbell}, {Lemen}, {Title}, {Boerner}, {Hurlburt}, {W{\"u}lser},
  {Mart{\'{\i}}nez-Sykora}, {Kleint}, {Golub}, {McKillop}, {Reeves}, {Saar},
  {Testa}, {Tian}, {Jaeggli}, \& {Kankelborg}}]{2014ApJ...792L..15P}
{Pereira}, T.~M.~D., {et~al.} 2014, \apjl, 792, L15

\bibitem[{{Rouppe van der Voort} {et~al.}(2009){Rouppe van der Voort},
  {Leenaarts}, {de Pontieu}, {Carlsson}, \& {Vissers}}]{2009ApJ...705..272R}
{Rouppe van der Voort}, L., {Leenaarts}, J., {de Pontieu}, B., {Carlsson}, M.,
  \& {Vissers}, G. 2009, \apj, 705, 272

\bibitem[{{Scharmer} {et~al.}(2003){Scharmer}, {Bjelksj{\"o}}, {Korhonen},
  {Lindberg}, \& {Petterson}}]{2003SPIE.4853..341S}
{Scharmer}, G.~B., {Bjelksj{\"o}}, K., {Korhonen}, T.~K., {Lindberg}, B., \&
  {Petterson}, B. 2003, in Society of Photo-Optical Instrumentation Engineers
  (SPIE) Conference Series, Vol. 4853, Society of Photo-Optical Instrumentation
  Engineers (SPIE) Conference Series, ed. S.~L. {Keil} \& S.~V. {Avakyan},
  341--350

\bibitem[{{Scharmer} {et~al.}(2008){Scharmer}, {Narayan}, {Hillberg}, {de la
  Cruz Rodr{\'{\i}}guez}, {L{\"o}fdahl}, {Kiselman}, {S{\"u}tterlin}, {van
  Noort}, \& {Lagg}}]{2008ApJ...689L..69S}
{Scharmer}, G.~B., {et~al.} 2008, \apjl, 689, L69

\bibitem[{{Sekse} {et~al.}(2012){Sekse}, {Rouppe van der Voort}, \& {De
  Pontieu}}]{2012ApJ...752..108S}
{Sekse}, D.~H., {Rouppe van der Voort}, L., \& {De Pontieu}, B. 2012, \apj,
  752, 108

\bibitem[{{Sekse} {et~al.}(2013{\natexlab{a}}){Sekse}, {Rouppe van der Voort},
  \& {De Pontieu}}]{2013ApJ...764..164S}
---. 2013{\natexlab{a}}, \apj, 764, 164

\bibitem[{{Sekse} {et~al.}(2013{\natexlab{b}}){Sekse}, {Rouppe van der Voort},
  {De Pontieu}, \& {Scullion}}]{2013ApJ...769...44S}
{Sekse}, D.~H., {Rouppe van der Voort}, L., {De Pontieu}, B., \& {Scullion}, E.
  2013{\natexlab{b}}, \apj, 769, 44

\bibitem[{{Tian} {et~al.}(2014){Tian}, {DeLuca}, {Cranmer}, {De Pontieu},
  {Peter}, {Mart{\'{\i}}nez-Sykora}, {Golub}, {McKillop}, {Reeves}, {Miralles},
  {McCauley}, {Saar}, {Testa}, {Weber}, {Murphy}, {Lemen}, {Title}, {Boerner},
  {Hurlburt}, {Tarbell}, {Wuelser}, {Kleint}, {Kankelborg}, {Jaeggli},
  {Carlsson}, {Hansteen}, \& {McIntosh}}]{tian2014science}
{Tian}, H., {et~al.} 2014, Science, 346

\bibitem[{{Tsiropoula} {et~al.}(2012){Tsiropoula}, {Tziotziou}, {Kontogiannis},
  {Madjarska}, {Doyle}, \& {Suematsu}}]{2012SSRv..tmp...65T}
{Tsiropoula}, G., {Tziotziou}, K., {Kontogiannis}, I., {Madjarska}, M.~S.,
  {Doyle}, J.~G., \& {Suematsu}, Y. 2012, \ssr, 65

\bibitem[{{van Noort} {et~al.}(2005){van Noort}, {Rouppe van der Voort}, \&
  {L{\"o}fdahl}}]{2005SoPh..228..191V}
{van Noort}, M., {Rouppe van der Voort}, L., \& {L{\"o}fdahl}, M.~G. 2005,
  \solphys, 228, 191

\bibitem[{{van Noort} \& {Rouppe van der Voort}(2008)}]{2008A&A...489..429V}
{van Noort}, M.~J., \& {Rouppe van der Voort}, L.~H.~M. 2008, \aap, 489, 429

\bibitem[{{Vissers} \& {Rouppe van der Voort}(2012)}]{2012ApJ...750...22V}
{Vissers}, G., \& {Rouppe van der Voort}, L. 2012, \apj, 750, 22

\bibitem[{{Yurchyshyn} {et~al.}(2013){Yurchyshyn}, {Abramenko}, \&
  {Goode}}]{2013ApJ...767...17Y}
{Yurchyshyn}, V., {Abramenko}, V., \& {Goode}, P. 2013, \apj, 767, 17

\end{thebibliography}

\end{document}